\documentclass[doublecol]{epl2}
\usepackage{graphicx}% Include figure files
\usepackage{dcolumn}% Align table columns on decimal point
\usepackage{bm}% bold math
\usepackage{epsf}
\usepackage{subfigure}
\usepackage{epstopdf}
\usepackage{amsmath}
\usepackage{amssymb}

\newcommand{\trace}{{\rm Tr}}

\begin{document}
\title{Second law and Landauer principle far from equilibrium}
\shorttitle{Second law and Landauer principle far from equilibrium}

\author{Massimiliano Esposito \inst{1} \and Christian Van den Broeck \inst{2}}
\institute{
 \inst{1} Center for Nonlinear Phenomena and Complex Systems, Universit\'e Libre de Bruxelles, CP 231, Campus Plaine, B-1050 Brussels, Belgium.\\
 \inst{2} Hasselt University, B-3590 Diepenbeek, Belgium.
}


\abstract{
The amount of work that is needed to change the state of a system in contact with a heat bath between specified initial and final {\it nonequilibrium} states is at least equal to the corresponding {\it equilibrium} free energy difference {\it plus} (resp. {\it minus}) temperature times the information of the final (resp. the initial) state relative to the corresponding equilibrium distributions.
}

\maketitle
%%%%%%%%%%%%%%%%%%%%%%%%%%%%%%%%%%%%%%%%%%%%%%%%%%%%%%%%%%%%%%%%%%%%%%
\section{Introduction}\label{intro}

Szilard was the first to realize that information processing, being a physical activity, has to obey the laws of thermodynamics \cite{Szilard1929}. In particular, he showed that the entropic cost for processing one bit of information is at least $k \ln 2$. The correct interpretation of this statement turns out to be rather subtle and the details (cost of measurement, of information storage and erasure, and of reversible and irreversible computation) have been the object of a longstanding and ongoing debate \cite{Brillouin53, 61Landauer, 82BennettThermoComp, Leff}. At the time of Szilard the transformation of information into work or vice-versa was a purely academic question. With the advent of high performance numerical simulations and the stunning developments in  nano- and bio-technology, the issue has received renewed attention \cite{Vedral, Leigh, BustamanteRitortPD, Horowitz10, HorowitzParrondo11, Fujitani10}. In particular, information to work transformation has been documented in computer simulations \cite{VandenBroeck07} and has been realized in several experiments \cite{Raizen, Leigh, Ciliberto05EPL, SagawaSano, BechingerBlickle}. Furthermore, spectacular developments in statistical mechanics and thermodynamics, including the work and fluctuation theorems \cite{Gallavotti95, Jarzynski97, Crooks00, HatanoSasa01, VdB06, AndrieuxGaspard07a, EspositoReview, BustamanteRitortPD} and the formulation of thermodynamics for single trajectories instead of ensemble averages \cite{Sekimoto98, Seifert05, EspositoVdBPRL10}, are very relevant in the context of information processing
\cite{Piechocinska00, Gaspard04b, Gaspard07Adv, VandenBroeck07, VdBNJP09, Segal10}. 
 
The second law of thermodynamics in the original formulation of macroscopic thermodynamics stipulates that the amount of work $W$, required to change the state of a system in contact with a heat bath between two different {\it equilibrium} states, is at least equal to the corresponding increase in {\it equilibrium} free energy $\Delta F^{\rm eq}$:
\begin{eqnarray}
 W - \Delta F^{\rm eq} \geq 0 \label{I1}.
\end{eqnarray} 
The equality sign is reached for a reversible transformation (the system remains at equilibrium all along the transformation). Note that $W$ refers to the work   performed on the system. In particular, work can be derived ($W<0$) only if the free energy of the system decreases. 

In this paper we give a straightforward and rather general thermodynamic proof, underpinned by exact arguments from statistical mechanics, that for an initial and final condition with distribution $\rho(0)$ and $\rho(t)$ different from the corresponding equilibrium distributions $\rho^{\rm eq}(0)$ and $\rho^{\rm eq}(t)$, an extra amount of work can be extracted or needs to be dispensed, namely:
\begin{eqnarray}
W_{\rm irr} \equiv W-\Delta F^{\rm eq} \geq  T \Delta I. \label{br} 
\end{eqnarray}
Here $W_{irr}$ is the so-called irreversible work and $\Delta I=I(t)-I(0)$ with
\begin{eqnarray}
I = D[\rho||\rho^{eq}]=  \trace \rho \ln \rho - \trace \rho \ln \rho^{\rm eq} \geq 0
\label{DefRelEntropy}
\end{eqnarray}
the relative entropy between the nonequilibrium and equilibrium distributions $\rho$ and $\rho^{\rm eq}$, respectively \cite{CoverThomas,Qian01b}. $I$ can also be identified  as the amount of information  that needs to be processed to switch from the known equilibrium distribution $\rho^{\rm eq}$ to the distribution $\rho$ under consideration \cite{CoverThomas}.
While this result has been recently derived for open Hamiltonian systems in \cite{Hasegawa10b,Hasegawa10}, we demonstrate in the present letter its greater generality by obtaining it from the nonequilibrium version of the second law of thermodynamics combined with the nonequilibrium Landauer principle. Furthermore, these results will be proven explicitly in the subsequent sections for stochastic Markovian dynamics, for isolated driven systems and for open driven systems.
We are using a quantum mechanical notation (with $\rho$ representing a density matrix and $\trace$ the trace), but the same result applies to classical systems (where $\rho$ stands for the probability density in state space and $\trace$ the integral over state space). 

%%%%%%%%%%%%%%%%%%%%%%%%%%%%%%%%%%%%%%%%%%%%%%%%%%%%%%%%%%%%%%%%%%%%%%
\section{Second law and Landauer principle}\label{mainres}

We consider a system described by a time-dependent Hamiltonian $H(t)$. As is usual in statistical mechanics, we characterize its state by a density matrix $\rho(t)$.  The system energy is the expectation value of the system Hamiltonian $H(t)$:
\begin{eqnarray}
E(t) = \trace \rho(t) H(t). \label{Energy}
\end{eqnarray}
We also introduce as {\it nonequilibrium} system entropy, the von Neumann (or Shannon) entropy:
\begin{eqnarray}
S(t) = - \trace \rho(t) \ln \rho(t) \label{Shannon},
\end{eqnarray}
and the corresponding {\it nonequilibrium} system free energy:
\begin{eqnarray}
F(t) = E(t) - T S(t) \label{FreeEnergy}.
\end{eqnarray}
Here $T$ is the temperature of an ideal heat bath with which the system is in contact. Furthermore the latter can exchange work
with an ideal (i.e., non-dissipative) work source. Let us call $W(t)$ and $Q(t)$ the work  performed on the system and the heat coming from the ideal heat bath after a time interval $t$. 
Following conservation of energy (first law of thermodynamics) the corresponding energy 
change $\Delta E(t)$ of the system is given by:
\begin{eqnarray}
\Delta E(t)=W(t)+Q(t). \label{1Law}
\end{eqnarray}

For the specific case of a reversible process, the system is at equilibrium with the bath at all times and hence is characterized by the ``instantaneous" equilibrium distribution $\rho^{\rm eq}(t)$, namely the canonical distribution corresponding to the ``instantaneous" Hamiltonian $H(t)$:
\begin{eqnarray}
\rho^{\rm eq}(t) \equiv \exp{\{-\beta \big( H(t) - F^{\rm eq}(t) \big) \}} \label{CanDM}.
\end{eqnarray}
By setting $\rho(t)=\rho^{\rm eq}(t)$  in  (\ref{Energy}) ,  (\ref{Shannon}) and (\ref{FreeEnergy}), we obtain the corresponding equilibrium values for energy, $E^{\rm eq}(t)$, entropy $S^{\rm eq}(t)$, and free energy $F^{\rm eq}(t)$.\\

%%%%%%%%%%%%%%%%%%%%%%%%
\begin{figure}[h]
\rotatebox{0}{\scalebox{0.85}{\includegraphics{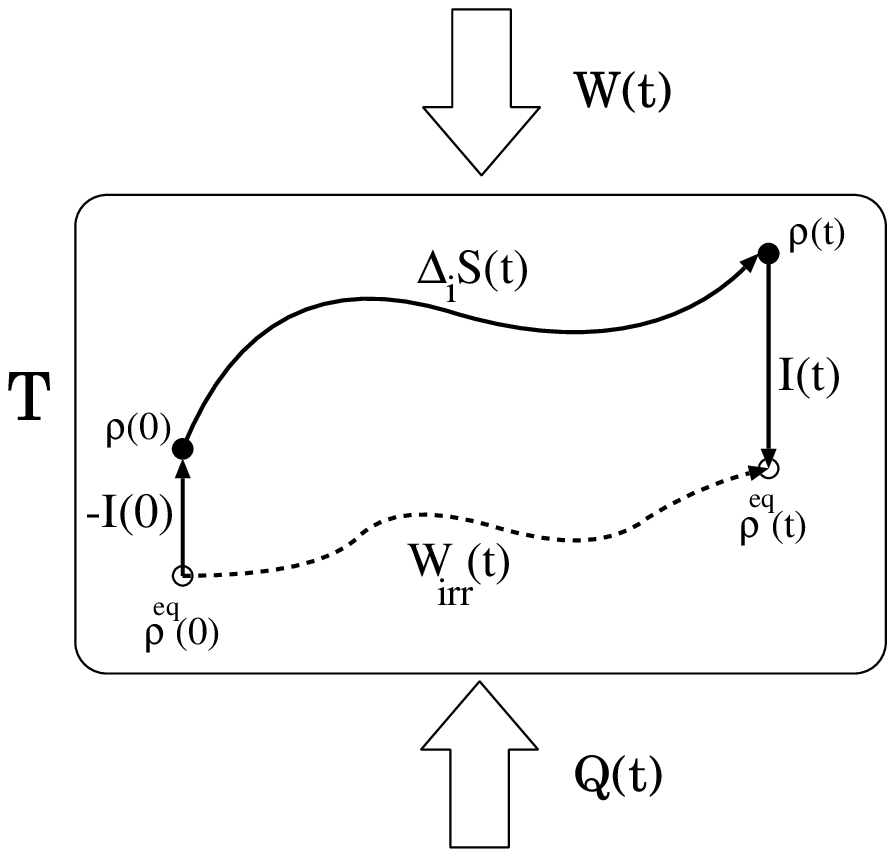}}}
\caption{Schematic illustration of equation (\ref{2LawBis}).}
\label{plot1}
\end{figure}
%%%%%%%%%%%%%%%%%%%%%%%%

We now state the two major results of this paper.\\

{\it Nonequilibrium second law}:
The change in the {\it nonequilibrium} system entropy $\Delta S$ consist of a reversible contribution due to the heat flow and called 
entropy flow $\Delta_{\rm e} S$, and of an irreversible (non-negative) contribution called entropy production $\Delta_{\rm i} S$: 
\begin{eqnarray}
&& \Delta S(t) = \Delta_{\rm i} S(t) + \Delta_{\rm e} S\label{2Lawtris} \\
&& \hspace {-0.3cm} \Delta_{\rm i} S(t) \ge 0 \ \ , \ \  \Delta_{\rm e} S=Q/T \nonumber
\end{eqnarray}

An equivalent formulation of the nonequilibrium second law can be readily obtained by combining 
the first law (\ref{1Law}) with the change in {\it nonequilibrium} free energy (\ref{FreeEnergy}):
\begin{eqnarray}
T \Delta_{\rm i} S(t) = W(t) - \Delta F(t) \geq 0 \label{2Law}.
\end{eqnarray}\\
In other words, the work that can be derived, $-W$, is at most equal to the decrease in {\it nonequilibrium} free energy $- \Delta F$.\\

{\it Nonequilibrium Landauer principle}:
The free energy of a nonequilibrium state is higher than that of the corresponding equilibrium state by an amount 
equal to the temperature times the information $I$ needed to specify the nonequilibrium state:
\begin{eqnarray}
F(t)-F^{\rm eq}(t)=T I(t)\equiv T D[\rho(t)||\rho^{\rm eq}(t)]  \geq 0. \label{landauer}
\end{eqnarray}
Combination of the above results allows to rewrite the second law under the form of the nonequilibrium Landauer principle 
\begin{eqnarray}
W_{\rm irr}(t) &\equiv& W(t)-\Delta F^{\rm eq}(t) \nonumber \\
&=& T \Delta_{\rm i} S(t) + T \Delta I(t) \label{2LawBis}
\end{eqnarray}
which leads to the result Eq.(\ref{br}) in the introduction.
Hence, contrary to the case of transitions between equilibrium states where $\Delta I(t)=0$ and thus $W_{\rm irr}(t)\geq 0$, 
the irreversible work can become negative if the reduction of the information $\Delta I(t) \leq 0$ is greater than the change 
in entropy production $\Delta_{\rm i} S \geq 0$.\\

The proof of (\ref{landauer}) is immediate, since $D\ge 0$. The derivation of the nonequilibrium 
second law (\ref{2Law}) will be given for three representative scenarios in the next section. \\

We first proceed with a further discussion of the above results.

1) Onsager, Prigogine, and others realized that the second law can be formulated for systems evolving in a state of local equilibrium. Entropy is then defined as a local quantity obeying a usual balance equation with $d_{\rm e} S$ representing the entropy flow and $d_{\rm i} S$ the irreversible local entropy production. The second law becomes a statement that can be applied locally in space and time $d_{\rm i} S /dt \geq 0$ \cite{Prigogine, GrootMazur}. The nonequilibrium second law (\ref{2Lawtris}) is a far from equilibrium generalization of this result. 

2) For an ideal reservoir which only exchanges entropy but does not irreversibly produce it, the change in the reservoir entropy is minus the entropy flow
\begin{eqnarray}
\Delta S_{\rm r}(t) = - Q(t)/T = -\Delta_{\rm e} S(t) \label{ResEnt}
\end{eqnarray}
As a result, the irreversible entropy production can be understood as the ``total" entropy change, $\Delta S_{\rm tot}$, i.e. the change in the system plus reservoir:
\begin{eqnarray}
\Delta_{\rm i} S(t) = \Delta S_{\rm tot} \equiv \Delta S(t) + \Delta S_{\rm r}(t) \geq 0 \label{2Law4}.
\end{eqnarray}

3) The optimal scenario with respect to work generation corresponds to a minimal most negative irreversible work $W^{\rm min}_{\rm irr}$, namely
\begin{eqnarray}
W_{\rm irr}(t) \geq W^{\rm min}_{\rm irr}(t) = - T D[\rho(0)||\rho^{\rm eq}(0)] \label{WorkIrrMin}
\end{eqnarray}
To reach this lower bound, two conditions need to be simultaneously satisfied. First, no information has to be left in the final distribution ($I(t)=0$), i.e., the final state of the system corresponds to equilibrium. Second, the transformation has to be done reversibly $\Delta_{\rm i} S(t)=0$. The lower bound can be reached by the following procedure (see also \cite{Hasegawa10b,Hasegawa10}). First perform a sudden quench at the initial time from the old Hamiltonian $H(0)$ to a new one $H^*(0)$, such that the original nonequilibrium initial condition becomes canonical equilibrium with respect to the new Hamiltonian (at the same temperature as the bath), $\rho(0)=\exp\{-\beta [H^*(0)-F^{{\rm eq},*}(0)]\}=\rho^{{\rm eq},*}(0)$. This is obviously a theoretical construction, serving as a proof of  principle rather than as a practical prescription. One may need a (infinitely) large number of  parameters controlling the location of the individual energy levels. The quench process does not change the entropy (since the distribution is not changed), but will require an average amount of irreversible work  equal to $W_{\rm irr}({\rm quench})= \trace \{\rho(0)[H^*(0)-H(0)]\}-\{F^{{\rm eq},*}(0)-F^{\rm eq}(0)\}=-T I(0)$. Next, starting from this new equilibrium state at time $0$, the final equilibrium state at time $t$ can be reached by a quasi-static change of the Hamiltonian - supposing $t$ is large enough to permit this - to its desired final form. The corresponding irreversible work as well as the entropy production vanish. The net result will be that $W_{\rm irr}(t) = W^{\rm min}_{\rm irr}(t) = - T I(0)$. 

4) We present a simple explicit illustration of how the lower bound $W^{\rm min}_{\rm irr}(t)$ can be reached. Consider an N-particle ideal gas enclosed in a box of volume $V$, in contact with a heat bath at temperature $T$. We initially confine the gas to the left volume $V/2$ and let it reach thermal equilibrium. At time $t=0$, the wall is removed and the former equilibrium state becomes a nonequilibrium initial condition $\rho(\Gamma)$ (with $\Gamma$ the  classical coordinates of all gasparticles in phase space). It is obviously zero in the right half of the volume and is larger in the left half than the equilibrium distribution in the full volume $\rho^{\rm eq}(\Gamma)$, by a factor $2^N$. Hence the relative entropy is given by (using the physically relevant limit $x \ln x \rightarrow 0$ for $x\rightarrow 0$) :
\begin{eqnarray}
I(0) 
%&=& \int \big( \prod_i d\Gamma_i \big) \; \rho_L(\{\Gamma_i\}) \ln \big( \rho_L(\{\Gamma_i\})/\rho(\{\Gamma_i\}) \big) \nonumber \\
&=&  \int d\Gamma \; \rho(\Gamma) \ln \big( \rho(\Gamma)/\rho^{eq}(\Gamma) \big) \nonumber \\
&=& \int d\Gamma \; \rho(\Gamma) \ln \big( 2^N \big)=N \ln 2.
\end{eqnarray}
If the gas is left to expand freely into the entire box until it reaches equilibrium, no work is extracted and the information stored in $T I(0)$ is completely lost into entropy production $\Delta_{\rm i} S(t)=I(0)$. However, if the wall is instantaneously reintroduced and the gas isothermally and reversibly expands against it, $T I(0)$ is completely converted into the extracted work $-W=\int_{V/2}^{V}dV P=N T \ln 2$ (since $P = T N/V$). We note that work and irreversible work are identical here because $\Delta F^{\rm eq}=0$.

5) If the system is assumed to be initially at equilibrium, we recover the conclusion of \cite{JarzynskiEPL09} that the irreversible work is always positive or zero,
\begin{eqnarray}
W_{\rm irr}(t) \geq T D[\rho(t)||\rho^{\rm eq}(t)] \geq 0 \label{JarzRes}
\end{eqnarray}
Our result (\ref{2Law}) is more general since both initial and final condition are arbitrary. In particular, the irreversible work can be negative (more work than the equilibrium free energy difference between the initial and final state can be extracted) if the system has been initially prepared in a nonequilibrium state. 
   
6) An obvious and also intriguing procedure to realize a distribution $\rho$ different from $\rho^{\rm eq}$ is by measuring the state of the system. A perfect measurement would yield a delta-function distribution. More realistically, one expects - due to the measurement error or quantum uncertainty - typically a broadened distribution around a most probable state. According to (\ref{br}), we are then able to extract an additional amount of work $T I$ (which could be quite substantial as the measurement becomes more precise). Alternatively, one can say that the measurement has decreased the total entropy by $I$. The resulting apparent violation of the second law is an example of the Szilard information-to-work conversion. In fact, the processing of the information $I$ by a physical device  will at least off-set the gain of work or decrease of total entropy that was realized in the measurement. 
Without entering in further details, we cite two recent works in which this information-to-work conversion is illustrated, namely one dealing with the  case of a Brownian particle in a manipulated potential \cite{SeifertEPL11} and the other with a Hamiltonian particle \cite{JarzynskiVaik11}. Note also that a related result was derived in a different context (quantum system subject to feedback control) in \cite{SagawaUedaPRL08,SagawaUedaPRL09}, and verified for the rectification of a Brownian particle moving in a staircase potential \cite{SagawaSano} (see also \cite{VdBNatPhys}).

7) In the case of a small system, one can raise the question of the role of the interaction energy \cite{JarzynskiReply04, Jarzynski07b, Peliti08, CampisiHanggiTalknerRev}. Any entropy production or energetic contribution pertaining to the interaction needs to be counted as part of the system energy, see also the discussion for open systems given below.

%%%%%%%%%%%%%%%%%%%%%%%%%%%%%%%%%%%%%%%%%%%%%%%%%%%%%%%%%%%%%%%%%%%%%%
\section{Stochastic system} \label{StochDyn}

We first derive the second law in the context of stochastic thermodynamics \cite{EspositoHarbola07PRE, EspoVdB10_Da, EspoVdB10_Db, SeifertST08}, 
which is applicable to both classical (Langevin or Master equation) or semi-classical systems (quantum Master equation). 
This procedure also provides a detailed and explicit expression for the various thermodynamic quantities involved. 
Let $\rho_i(t)$ be the probability to find the system in state $i$ with energy $H_i(t)$ at time $t$. 
We assume a Markovian dynamics so that its time evolution is governed by the following Master equation: 
\begin{eqnarray}
\dot{\rho}_i(t) = \sum_j M_{ij}(t) \rho_j(t). \label{ME}
\end{eqnarray}
$M_{ij}$ is the transition rate matrix, with $\sum_i M_{ij}(t)=0$. 
The time dependence of the states energy $H_i(t)$ is a result of the interaction with the external work source.
The basic physical ingredient for stochastic thermodynamics is the requirement that the rates  reproduce the proper equilibrium state and satisfy local detailed balance. For the simplest case considered here (heat exchange with a single bath), one has:
\begin{eqnarray}
\ln \frac{M_{i,j}(t)}{M_{j,i}(t)} = - \beta [H_i(t)-H_j(t)]. \label{LDB}
\end{eqnarray}
Here we have assumed that the instantaneous stationary solution of the master equation (i.e., the eigenvector 
with zero eigenvalue of $W(t)$) corresponds to the instantaneous canonical equilibrium probability (\ref{CanDM}), 
which is reached when the time-dependence is frozen. 
It is now straightforward to explicitly verify that the first law (\ref{1Law}) as well as the nonequilibrium 
version of the second law of thermodynamics (\ref{2Law}) hold true (for details see e.g. \cite{EspoVdB10_Da}). 
The energy and entropy are respectively given by
\begin{eqnarray}
E(t) = \sum_i \rho_i(t) H_i(t)  \; , \; S(t) = -\sum_i \rho_i(t) \ln \rho_i(t) ,\label{StochEntr} 
\end{eqnarray}
heat and work by
\begin{eqnarray}
\dot{Q}(t) = \sum_i \dot{\rho}_i(t) H_i(t) \; , \; \dot{W}(t) = \sum_i \rho_i(t) \dot{H}_i(t) . \label{StochWork} 
\end{eqnarray}
The Master equation (\ref{ME}) with (\ref{LDB}) leads to the familiar entropy balance equation:
\begin{eqnarray}
\dot{S}(t)=\dot{S}_{\rm i}(t) +\dot{S}_{\rm e}(t) \label{StochEPbis}
\end{eqnarray}
with entropy flow rate $\dot{S}_{\rm e}(t)=\beta \dot{Q}(t)$ and irreversible entropy production rate given by:
\begin{eqnarray}
\dot{S}_{\rm i}(t)  = \sum_{i,j} M_{ij}(t) \rho_j(t) \ln \frac{M_{ij}(t) \rho_j(t)}{M_{ji}(t) \rho_i(t)} \geq 0. \label{StochEP}
\end{eqnarray}
There are no assumptions about initial and final states, nor on the type of time-dependence of the rates (other than the fact that the Markovian approximation and local equilibrium condition remain valid). In fact, not only the positivity of the change in entropy production is thus proven, but the stronger 
condition that the rate of entropy production is positive. Hence our main result (\ref{2LawBis}) 
can be replaced by:
\begin{eqnarray}
\dot{W}_{\rm irr}(t) = \dot{W}(t)-\dot{F}^{\rm eq}(t)= T \dot{S}_{\rm i}(t) + T \dot{I}(t) \label{2LawBisStoch}
\end{eqnarray}
Beside reproducing the first and second law, stochastic thermodynamics also satisfies the zeroth law:
in absence of driving (i.e. time-independent energies $H_i$), the system relaxes to the corresponding 
equilibrium (\ref{CanDM}). As a consequence, for slow changes of the states energies $H_i(t)$ (i.e. for 
reversible transformations), the probability distribution is the instantaneous equilibrium probability 
(\ref{CanDM}) which satisfies detailed balance at all times. As expected, reversible 
transformations give rise to zero entropy production (\ref{StochEP}). Faster transformations will 
generate positive entropy production. Since stochastic thermodynamics gives explicit expressions 
for all quantities involved, the irreversible work can be calculated for specific models or under 
specific conditions. In particular, the issue of extracting maximum work in finite time has been 
discussed in great detail \cite{Seifert07b, 08SeifertJCP, EspKawLindVdBEPL10, EspoKawLindVdB_PRE_10, 
10EspoKawLindVDB_PRL, SeifertEPL11}.

It is worth noting that for stochastic dynamics, (\ref{2LawBis}) or (\ref{2LawBisStoch}) can be seen as a special case of an even more general version of the second law, valid for systems in contact with multiple baths, which states that entropy production $\dot{S}_i$ is the sum of an adiabatic and a nonadiabatic contribution, see eq.(21) of \cite{EspositoHarbola07PRE} (see also \cite{Harris07, EspoVdB10_Da, Ge09, GeQian10}). Indeed, because we consider here a single reservoir, the adiabatic contribution vanishes and the boundary and the driving term, whose sum is the nonadiabatic contribution, see eq.(16) of \cite{EspositoHarbola07PRE}, become respectively $-\dot{I}$ and $\dot{W}_{\rm irr}/T$.

%%%%%%%%%%%%%%%%%%%%%%%%%%%%%%%%%%%%%%%%%%%%%%%%%%%%%%%%%%%%%%%%%%%%%%
\section{Isolated driven system}\label{drivenHam}

We consider a system, described by Hamiltonian dynamics under influence of a time dependent Hamiltonian $H(t)$, initially in state $\rho(0)$. The Hamiltonian evolution of the system density matrix $\rho(t)$ implies that the von Neumann 
entropy of the system $S(t)$, given by (\ref{Shannon}), is invariant in time $\Delta S(t)=0$.
The corresponding change in nonequilibrium free energy $F=E-TS$ is therefore equal to the energy change $\Delta F(t)=\Delta E(t)$.
Since there is no heat exchange $Q(t)=0$, the latter is equal to the work (first law (\ref{1Law})), resulting from the time dependence of the Hamiltonian: 
\begin{eqnarray}
W(t) = \Delta E(t) = \trace \rho(t) H(t) - \trace \rho(0) H(0) .\label{IsolatedWork}
\end{eqnarray}
No heat also implies a zero entropy flow $\Delta_{\rm e}S(t)=0$ which, using (\ref{2Law}), leads to a zero entropy production $\Delta_{\rm i}S(t)=0$.

Consequently, a short calculation shows that the irreversible work, defined relative to an ideal bath at temperature $T$, is given by:
\begin{eqnarray}
W_{\rm irr}(t) = \Delta I(t) = T D[\rho(t)||\rho^{\rm eq}(t)] - T D[\rho(0)||\rho^{\rm eq}(0)] \label{CentralIsolated} 
\end{eqnarray}
which is the nonequilibrium Landauer principle (\ref{2LawBis}) where $\Delta_{\rm i} S(t)=0$.
For equilibrium initial conditions, this reduces to the result from \cite{VandenBroeck07,VdBNJP09,JarzynskiEPL09},
\begin{eqnarray}
W_{\rm irr}(t) = T D[\rho(t)||\rho^{\rm eq}(t)] \geq 0. \label{JarzRes}
\end{eqnarray}

To provide a meaning to the temperature $T$ in (\ref{CentralIsolated}), we connect the system at the final time with the ideal bath. 
The latter will relax from $\rho(t)$ to $\rho^{\rm eq}(t)$ (the zeroth law) thus inducing an entropy change $\Delta S(t)=S^{\rm eq}(t)-S(0)$ and an 
heat flow $Q(t)=\Delta E(t)=E^{\rm eq}(t)-E(t)$. As a result the nonequilibrium second law (\ref{2Lawtris}) becomes
\begin{eqnarray}
T \Delta_{\rm i} S(t) &=& F^{\rm eq}(t)-(E(t)-T S(0)) \nonumber \\
&=& T D[\rho(t)||\rho^{\rm eq}(t)] \geq 0 \label{EPisol}
\end{eqnarray}
Using (\ref{CentralIsolated}), this implies that 
\begin{eqnarray}
W_{\rm irr}(t) = T \Delta_{\rm i} S(t) - T D[\rho(0)||\rho^{\rm eq}(0)] \label{CentralIsolated2}
\end{eqnarray}
which is the nonequilibrium Landauer principle (\ref{2LawBis}) where $I(t)=0$ due to the fact 
that the final state of the system is assumed to be at equilibrium $\rho(t)=\rho^{\rm eq}(t)$.
It is interesting to compare the form of the irreversible work before and after the connection with the ideal reservoir, i.e. (\ref{CentralIsolated}) with (\ref{CentralIsolated2}). 
We observe that the information $I(t)$ remaining in the final state $\rho(t)$ (i.e. the part of the initial information $I(0)$ which has not been converted into work during the isolated dynamics) 
has been converted into entropy production (\ref{EPisol}) after the relaxation with the ideal bath has occurred.

%%%%%%%%%%%%%%%%%%%%%%%%%%%%%%%%%%%%%%%%%%%%%%%%%%%%%%%%%%%%%%%%%%%%%%
\section{Open driven system}\label{Open}

We consider now that the system $S$ with Hamiltonian $H(\tau)$ is open, i.e. coupled to a finite 
bath $B$ with Hamiltonian $H_B$ ($\tau$ denotes the time which varies between $0$ and $t$).
The Hamiltonian of the total (isolated) system reads 
\begin{eqnarray}
H_{tot}(\tau) = H(\tau) + H_B + V(\tau) \label{HamiltonianDef},
\end{eqnarray}
where $V(\tau)$ describes the interaction between the system and the bath.  
The system entropy is given by the von Neumann entropy (\ref{Shannon}), $S(t) = - \trace \rho(t) \ln \rho(t)$, where the system density matrix is 
the reduced density matrix obtained by tracing the full density matrix over the bath degrees of freedom $\rho(\tau)=\trace_B \rho_{tot}(\tau)$.
The work, heat and energy are respectively given by
\begin{eqnarray}
W(\tau) &=& \trace \rho_{tot}(\tau) H_{tot}(\tau) - \trace \rho_{tot}(0) H_{tot}(0) \label{OpenWork}\\
Q(\tau) &=& \trace \rho_{tot}(0) H_{B} - \trace \rho_{tot}(\tau) H_{B} \label{OpenHeat}\\
E(\tau) &=& \trace \rho_{tot}(\tau) [H(\tau)+V(\tau)] \label{OpenEnergy}
\end{eqnarray}
We easily verify that these definitions satisfy the first law (\ref{1Law}).
The nonequilibrium free energy is $F(\tau)=E(\tau)-T S(\tau)$. 
We also note that the interaction term in the total Hamiltonian is included 
in the system energy, while it does not enter in the definition of heat. 

In order to make connection with our main results, we make the following two assumptions:

{\it Assumption I}: The system and the bath are initially uncorrelated. 
The system in an arbitrary state $\rho(0)$ and the bath at equilibrium, i.e.
\begin{eqnarray}
\rho_{tot}(0) = \rho(0) \rho_B^{\rm eq}. \label{Assum1} 
\end{eqnarray}
It has been shown in \cite{EspoLindVdBNJP10} (see also \cite{Jarzynski99}) that under this single assumption, 
the second law (\ref{2Lawtris}) or (\ref{2Law}) is satisfied (and positivity is proven):
\begin{eqnarray}
\Delta_{\rm i} S(t) = \frac{W(t) - \Delta F(t)}{T} = \Delta S(t) - \frac{Q(t)}{T} \geq 0 \label{2LawQO}.
\end{eqnarray}

{\it Assumption II}: At the beginning and at the end of the process, the 
interaction between the system and the bath is, respectively, turned on and off,
\begin{eqnarray}
V(0) = V(t) = 0. \label{Assum2}
\end{eqnarray}
This condition allows the system state functions such as energy and free energy to be expressed 
exclusively in terms of system quantities without the contribution from the system-bath coupling 
at the beginning and at the end of the process. This assumption is essential to make the energy 
definition (\ref{OpenEnergy}) compatible with definition (\ref{Energy}), else they differ by the 
system-bath coupling. The same is therefore automatically true for the free energy which means 
that (\ref{landauer}) is satisfied at final time $t$ and at initial time $0$
\begin{eqnarray}
&&\hspace{-0.5cm} F(t)-F^{\rm eq}(t)=T I(t)\equiv T D[\rho(t)||\rho^{\rm eq}(t)]  \geq 0  \nonumber \\
&&\hspace{-0.5cm} F(0)-F^{\rm eq}(0)=T I(0)\equiv T D[\rho(0)||\rho^{\rm eq}(0)]  \geq 0. \label{landauerQO}
\end{eqnarray}

Combining the second law (\ref{2LawQO}) [valid thanks to Assumption I] with (\ref{landauerQO}) 
[valid thanks to Assumption II], we easily recover our central result (\ref{2LawBis}):
\begin{eqnarray}
W_{\rm irr}(t) &\equiv& W(t)-\Delta F^{\rm eq}(t) \nonumber \\
&=& T \Delta_{\rm i} S(t) + T \Delta I(t) \label{2LawBisQO}
\end{eqnarray}
This result is valid for arbitrary coupling strength $V(\tau)$ between time $0$ and $t$. 
In the weak coupling regime where the contributions from the interaction strength can be 
neglected, (\ref{2LawBisQO}) can been derived without assumption II \cite{Lutz11}.

%%%%%%%%%%%%%%%%%%%%%%%%%%%%%%%%%%%%%%%%%%%%%%%%%%%%%%%%%%%%%%%%%%%%%%
\section{Conclusions}\label{conclusions}

Our main result (\ref{2LawBis}) has been derived for three types of basic dynamics:
stochastic dynamics, Hamiltonian dynamics of an isolated driven system and Hamiltonian 
dynamics of an open driven system. This result shows that the irreversible work consist 
of two contributions: a non-negative entropy production contribution and a boundary term 
containing the information stored in the initial and final condition of the system 
probability distribution. The latter can be negative and even lead to negative 
values for the irreversible work. The above result vindicates the view on the second law 
pioneered by Ilya Prigogine: the entropy production is the basic non-negative quantity.

%%%%%%%%%%%%%%%%%%%%%%%%%%%%%%%%%%%%%%%%%%%%%%%%%%%%%%%%%%%%%%%%%%%%%%
\acknowledgments

M. E. is supported by the Belgian Federal Government (IAP project ``NOSY") and by the 
European Union Seventh Framework Programme (FP7/2007-2013) under grant agreement 256251.

%%%%%%%%%%%%%%%%%%%%%%%%%%%%%%%%%%%%%%%%%%%%%%%%%%%%%%%%%%%%%%%%%%%%%%

%%%%%%%%%%%%%%%%%%%%%%%%%%%%%%%%%%%%%%%%%%%%%%%%%%%%%%%%%%%%%%%%%%%%%%
\end{document}